\pdfoutput=1
\documentclass[structabstract]{aa}  
\usepackage{graphicx}
\usepackage{natbib}
\usepackage{subfig}
\usepackage{epsfig}
\usepackage{ifpdf} 
\usepackage{fancyvrb}

\usepackage{txfonts}
\begin{document}
   \title{Neutron star masses from hydrodynamical effects in obscured sgHMXBs}


   \author{A. Manousakis 
          \inst{1,2}
          \and
          R. Walter \inst{1,2}
          \and
          J. M. Blondin \inst{3}
          }

   \institute{ ISDC Data Center for Astrophysics, Universit\'e de Gen\`eve, Chemin d'Ecogia 16, CH-1290 Versoix, Switzerland \\
             \email{Antonios.Manousakis@unige.ch}
        \and Observatoire de Gen\`eve, Universit\'e de Gen\`eve,  Chemin des Maillettes 51, CH-1290 Versoix, Switzerland        
                \and
Department of Physics, North Carolina State University, Raleigh, NC 27695-8202, USA\\        
             }

   \date{Received May 30, 2012; accepted Sept. 05, 2012}

 
  \abstract
   {A  population of  obscured supergiant High Mass X-ray Binaries (sgHMXBs) has been discovered by INTEGRAL. 
  X-ray wind tomography of IGR J17252-3616 inferred a slow wind  velocity to account for the enhanced obscuration.  }
   {The main goal of this study is to understand under which conditions high obscuration could occur.}
   {We have used an hydrodynamical code to simulate the flow of the stellar wind around the neutron star. A grid of simulations
   was used to study the dependency of the  absorbing column density and of the X-ray light-curves on the model 
   parameters. A comparison between 
   the simulation results and the observations  of IGR J17252-3616 provides an estimate on these parameters. }
   {We have constrained  the wind terminal  velocity to  500-600 km s$^{-1}$ and 
    the neutron star mass  to $1.75 - 2.15\,  M_{\odot}$.} 
   {We have confirmed that the initial hypothesis of a slow wind velocity with a moderate mass loss rate is valid.  
   The mass of the neutron star can be constrained by studying its impact on the accretion flow.}



   \maketitle
%

\section{Introduction}

In  classical supergiant High Mass X-ray Binaries (sgHMXBs)  neutron stars are orbiting at a distance of $\alpha\sim 1.5-2$ R$_{*}$ from their companion stars. 
The donors, in these systems,  are  OB supergiants  with mass loss rates
 of the order of $\sim10^{-6}\, M_{\odot}$ yr$^{-1}$ and  wind terminal velocities of $\sim$ 1500 km s$^{-1}$.

\citet{Blondin90,Blondin91} modeled  the interactions between the neutron star 
and the stellar wind in Vela X-1  
and revealed that the wind of the massive star is disrupted by the gravity and photoionization of  the neutron star.
The imprint of these two parameters can be observed in the variability of the absorbing column density with orbital phase.
High-resolution soft X-ray spectroscopy of the brightest sources revealed a number of lines in emission, 
 constraining the ionization level of the gas \citep{watanabe}.

The  heavily obscured sgHMXBs \citep{Walter_et_al06} share some of the  characteristics of the classical sgHMXBs. 
The main difference between classical and obscured sgHMXBs is that the latter ones are much 
more absorbed in the X-rays. 
The absorbing column density ($N_{H} \ga 10^{23}$ cm$^{-2}$) is, on average, 10 times larger than in  classical systems 
and well above the galactic absorption   in the direction of the sources.

The obscured sgHMXB IGR $J17252-3616$ \citep[=EXO 1722-363;][]{Walter_et_al06,Zurita_et_al06,Thompson_et_al07} 
 is an eclipsing binary hosting a  pulsar with P$_{s}\sim$ 414 sec, an orbital period of P$_{o}\sim$ 9.74 days, and an orbital  radius of  $\alpha\approx$1.75 $R_{*}$. 
Ground-based observations \citep{Mason_et_al09,Mason_et_al09b}  showed that the donor star is likely a B0-5I or B0-1 Ia. 
Optical and IR observations confirmed the supergiant nature and  showed prominent P-Cygni profile \citep{Chaty_et_al08}

To explain the $XMM-Newton$ and $INTEGRAL$ observations, \citet[][hereafter MW11]{Manousakis+11} suggested  that the wind terminal velocity of the system is relatively, low, of the order  of $\upsilon_{\infty}\sim 500$ km s$^{-1}$. 
An ad-hoc modeling of a hydrodynamic trailing tail allowed  to reproduce the observed column density profile
 and the  observed iron Fe K$\alpha$ line emissivity.

In the present paper, we have used the hydrodynamic code VH-1 to study the wind dynamics of IGR J17252-3616 
under the assumption of a slow wind velocity. 
We start with a description of the hydrodynamic code and of the grid of  simulations in section \ref{sec:code}, the 
results are presented in section \ref{sec:results} and discussed in  
section \ref{sec:discussion}, followed by the conclusions in section \ref{sec:conclude}.


\section{Hydrodynamical Simulations} 
\label{sec:code}

 \subsection{The hydrodynamical code}

The full description of the VH-1\footnote{http://wonka.physics.ncsu.edu/pub/VH-1/} hydrodynamical code is given in \citet{Blondin90,Blondin91}. 
To simulate sgHMXB the code accounts for: i) the gravity of 
the primary and of the neutron star, ii) the radiative acceleration of the stellar wind of the primary star, and iii) 
the suppression of the stellar wind acceleration in the Str\"{o}mgren sphere of the neutron star.
The simulations take place  in the orbital plane, reducing the problem  to two dimensions.
The code uses the piecewise parabolic method developed by \citet{PPMCW}.

A computational grid of 600 radial by 247 angular zones, extending  from 1 to $\sim$ 15 R$_{*}$ and in 
angle from $-\pi$ to $+\pi$, has been employed. The grid structure is non-uniform, with cells size decreasing towards the neutron star.
The maximal resolution, reached close to the neutron star, is $\delta R\sim 10^{10}$ cm $\sim r_{acc}/3$, 
where $r_{acc}=2GM_{X}/(\upsilon_{orb}^{2}+\upsilon_{x}^2)$ is the accretion radius \citep{BondiHoyle}. 

The neutron star in IGR J17252-3616 orbits in a almost circular ($e<0.15$) and highly 
inclined ($i>80^{\circ}$) orbit \citep{Manousakis+11}. 
Therefore, a circular  ($e=0$) and  edge-on ($i=90^{\circ}$) orbit is assumed for simplicity. 
As most of the absorption comes from the accretion stream, close to the neutron star 
(within  $10^{12}$ cm), the variability of the absorbing column 
density with orbital phase does  not depend on
 small variations of the inclination angle.

The code produces  density and ionization \citep[$\xi= L_{X}/n r_{ns}^2$, 
 where $L_{X}$ is the average X-ray luminosity, 
 $n$ is the number density at the distance $r_{ns}$
 from the neutron star;][]{1969ApJ...156..943T} maps that are stored. 
These allow to determine the  simulated column density. As short time-scale variations occurs, we have 
calculated the time-averaged  orbital phase resolved column density and its 3$\sigma$  variability. 
The 3-D instantaneous mass accretion rate ($\dot{M}_{acc}$) onto the neutron star is also recorded by 
 translating the 2-D instantaneous mass accretion rate   along the rim of the accretion radius, $r_{acc}$. 
 From the mass accretion rate, we can infer the instantaneous X-ray luminosity of the neutron star. 	
For each simulations, the code was  run for about 6 to 8 orbits, corresponding to about 60 $-$ 80 days. This is enough 
for the wind to reach a stable configuration. The relaxation time is of the order of 0.8 orbits. The first 10 days of the simulations are therefore excluded from our analysis.

Although 3D hydrodynamic simulations are poorly studied in HMXB, 
differences between 2D and 3D  are expected.
In general, 3D simulations increase the instabilities and fluctuations.
These effects have been  studied in the idealized case of Hoyle-Lyttleton accretion \citep{1988ApJ...327L..73T,1999A&A...346..861R,2012ApJ...752...30B} and in  accreting pulsars where a formation of a planar accretion disk is expected from 2D codes, whilst accretion in 3D is unstable, allowing  the formation of filamentary structures \citep{1994ApJ...427..351R}.

\subsection{Stellar wind acceleration}

The winds of hot massive stars are characterized, observationally, by 
  the wind terminal velocity and  mass-loss rate. 
 The velocity  is  described by the  $\beta$-velocity law,
$ \upsilon=\upsilon_{\infty}(1-R_{*}/r)^{\,\beta}$
 where $\upsilon_{\infty}$ is the terminal velocity and 
 $\beta$ is the gradient of the velocity field. 
For supergiant stars,  values for wind terminal velocities 
and mass-loss rates are in the range $\upsilon_{\infty}\sim500-3000$ km s$^{-1}$ and
$\dot{M}_{\rm w}\sim 10^{-7} - 10^{-5}\, M_{\odot}$ yr$^{-1}$, respectively \citep{winds_from_hot_stars}.  
The donor star typically weights M$_{*}\ga 10\, M_{\odot}$ with a radius of 
R$_{*}\ga 10^{12}$ cm.

 The  winds  of massive supergiant stars are  radiatively driven by absorbing photons from the photosphere  \citep[CAK hereafter]{CAKwind}. 
The  modeling of  stellar wind is complicated by the fact the 
 radiation force is known to be unstable \citep{1984ApJ...284..337O}. 
   We use the CAK/Sobolev approximations with the finite disk correction \citep{1986ApJ...311..701F}
  which produces a smooth stellar wind with $\beta=0.8$.   
 However, regions in the stellar wind can differ from the predictions of the  CAK/Sobolev approximation when  growth of instabilities are taken into account \citep{Owocki+88}.

 \subsection{Photoionization}
 
 The radiative acceleration of the wind from the donor star primarily occurs due to the 
 UV line transitions that are suppressed by X-ray ionization.
  The effects of X-ray ionization on the radiative acceleration force are complicated due to the large
 number of ions and line transitions which contribute to the UV opacity \citep{1990ApJ...365..321S}.
 A critical ionization parameter can be defined, above which the radiative force 
is negligible. 
 For $\xi>10^{2.5}$ erg cm sec$^{-1}$, most of the elements  responsible for the wind acceleration are 
 fully ionized and,  therefore, the radiatively acceleration force vanishes. 
 At that level, light atoms (e.g. H) are fully ionized and do not contribute to the absorbing column density 
 while heavier atoms (e.g. C, N, O) are highly ionized \citep{Kallman82}
 
The main effect of the ionization is the 
 enhancement of the mass accretion rate onto the compact object.
The acceleration cutoff also triggers the formation of a dense wake  at the rim of the Str\"{o}mgren zone \citep{FarnssonFabian1980}, and  has an impact on  the absorption 
at late orbital phases. X-ray ionization can also affect the thermal state of the wind through X-ray heating and radiative cooling,  which
can create additional filamentary structures in the wind, with slightly different time-averaged absorbing column density \citep{Blondin90}.

\subsection{Parametrization}

In order to simulate IGR J17252-3616, we proceed in 
two steps. We started by investigating the parametrization of  the stellar wind of the donor star. 
This is  achieved using a 1-D simulation assuming spherical symmetry. 
Spectral classification allowed to constrain the donor star parameters \citep{Chaty_et_al08,Mason_et_al09,Mason_et_al09b}.
In our analysis we have adopted an effective temperature of $T_{eff}=30 000$ K, 
a stellar luminosity of $L_{*}=4\times 10^5\, L_{\odot}$, and a stellar radius of  $R_{*}=29\, R_{\odot}$.
The mass of the donor star is assumed to be $M_{*}=15\, M_{\odot}$ \citep{Takeuchi_et_al90,Thompson_et_al07}.

\begin{table*}
\caption{ Parameters of the 2-D simulations.
Three different groups are considered: \emph{Top:} Comparison between slow and normal wind terminal velocity for a number 
of mass-loss rates and different neutron star masses. \emph{Middle:} Simulations to study the dependency, of the most likely model, on the neutron star mass.
\emph{Bottom:} Fine tuning  of the neutron star mass and binary separation.   
}
\centering                          
\begin{tabular*}{0.9\textwidth}{@{\extracolsep{\fill}}  l  c c  c  c  c c c c }
\hline
\hline                 
\\
Model name$^{a}$                                      & group &  CAK-$\alpha^{b}$  & CAK-$k^{b}$    & $\rho_{0}^{b} $      & $\dot{M}_{\rm w}^{c}$ & $ \upsilon_{\infty} ^{d}$ & $\alpha^{e}$    &   $M_{X}^{f}$   \\
\\
\hline
\\
\verb=v12_ML2_A175_MNXYZ=   &  1 &    0.55                  & 0.05             &$1\cdot10^{-12}$               &  2                             &    1200                    &   1.75         &  1.55, 1.75, 1.95   \\

\verb=v12_ML10_A175_MNXYZ=   &  1 &0.55                  & 0.15             &$5\cdot10^{-12}$   &  10                            &    1200                    &   1.75         &  1.55, 1.75, 1.95  \\

\verb=v12_ML50_A175_MNXYZ=   &  1 & 0.55                  & 0.30             &$1\cdot10^{-11}$   &  50                            &    1200                    &   1.75         &  1.55, 1.75, 1.95    \\

\verb=v5.0_ML2_A175_MNXYZ=   &  1 & 0.35                  & 0.20             &$5\cdot10^{-12}$   &  2                             &    500                    &   1.75         &  1.55, 1.75, 1.95  \\

\verb=v5.0_ML10_A175_MNXYZ=   &  1& 0.35                  & 0.30             &$5\cdot10^{-11}$   &  10                             &    500                    &   1.75         &  1.55, 1.75, 1.95  \\

\verb=v5.0_ML50_A175_MNXYZ=   & 1 &  0.35                  & 0.55             &$1\cdot10^{-10}$   &  50                             &    500                    &   1.75         &  1.55, 1.75, 1.95   \\
\\
\hline
\\
\verb=v5.0_ML10_A175_MNXYZ=   &  2 & 0.35                  & 0.30             &$5\cdot10^{-11}$   &  10                             &    500                    &   1.75         &  1.5, 1.6,  1.7, 1.8, 1.9, 2.0  \\
\\
\hline
\\
\verb=v5.5_ML10_A173_MNXYZ=   & 3&  0.37                  & 0.28             &$5\cdot10^{-11}$   &  10                             &    550                    &   1.73         &  1.85, 1.90, 1.95, 2.00  \\
\verb=v5.5_ML10_A174_MNXYZ=   & 3& 0.37                  & 0.28             &$5\cdot10^{-11}$   &  10                             &    550                    &   1.74          &  1.85, 1.90, 1.95, 2.00  \\
\verb=v5.5_ML10_A175_MNXYZ=   &  3&0.37                  & 0.28             &$5\cdot10^{-11}$   &  10                             &    550                    &   1.75         &  1.85, 1.90, 1.95, 2.00  \\
\verb=v5.5_ML10_A176_MNXYZ=   &  3&0.37                  & 0.28             &$5\cdot10^{-11}$   &  10                             &    550                    &   1.76         &  1.85, 1.90, 1.95, 2.00  \\
\verb=v5.5_ML10_A177_MNXYZ=   &  3&0.37                  & 0.28             &$5\cdot10^{-11}$   &  10                             &    550                    &   1.77         &  1.85, 1.90, 1.95, 2.00  \\
\\
\verb=v6.0_ML10_A173_MNXYZ=   &  3&0.38                  & 0.27             &$3\cdot10^{-11}$   &  10                             &    600                    &   1.73         &  1.85, 1.90, 1.95, 2.00  \\
\verb=v6.0_ML10_A174_MNXYZ=   &  3&0.38                  & 0.27             &$3\cdot10^{-11}$   &  10                             &    600                    &   1.74          &  1.85, 1.90, 1.95, 2.00  \\
\verb=v6.0_ML10_A175_MNXYZ=   &  3&0.38                 & 0.27            &$3\cdot10^{-11}$   &  10                             &    600                    &   1.75         &  1.85, 1.90, 1.95, 2.00  \\
\verb=v6.0_ML10_A176_MNXYZ=   &  3&0.38                  & 0.27             &$3\cdot10^{-11}$   &  10                             &    600                    &   1.76         &  1.85, 1.90, 1.95, 2.00  \\
\verb=v6.0_ML10_A177_MNXYZ=   &  3&0.38                  & 0.27             &$3\cdot10^{-11}$   &  10                             &    600                    &   1.77         &  1.85, 1.90, 1.95, 2.00  \\
\\
\verb=v6.5_ML10_A173_MNXYZ=   &  3&0.395                  & 0.25             &$3\cdot10^{-11}$   &  10                             &    650                    &   1.73         &  1.85, 1.90, 1.95, 2.00  \\
\verb=v6.5_ML10_A174_MNXYZ=   &  3&0.395                & 0.25             &$3\cdot10^{-11}$   &  10                             &    650                    &   1.74          &  1.85, 1.90, 1.95, 2.00  \\
\verb=v6.5_ML10_A175_MNXYZ=   &  3&0.395                  & 0.25             &$3\cdot10^{-11}$   &  10                             &    650                    &   1.75         &  1.85, 1.90, 1.95, 2.00  \\
\verb=v6.5_ML10_A176_MNXYZ=   &  3&0.395                  & 0.25             &$3\cdot10^{-11}$   &  10                             &    650                    &   1.76         &  1.85, 1.90, 1.95, 2.00  \\
\verb=v6.5_ML10_A177_MNXYZ=   &  3&0.395                  & 0.25             &$3\cdot10^{-11}$   &  10                             &    650                    &   1.77         &  1.85, 1.90, 1.95, 2.00  \\
\\
\verb=v7.0_ML10_A173_MNXYZ=   &  3&0.417                  & 0.22             &$2\cdot10^{-11}$   &  10                             &    700                    &   1.73         &  1.85, 1.90, 1.95, 2.00  \\
\verb=v7.0_ML10_A174_MNXYZ=   &3&  0.417                  & 0.22             &$2\cdot10^{-11}$   &  10                             &    700                    &   1.74          &  1.85, 1.90, 1.95, 2.00  \\
\verb=v7.0_ML10_A175_MNXYZ=   & 3& 0.417                  & 0.22             &$2\cdot10^{-11}$   &  10                             &    700                    &   1.75         &  1.85, 1.90, 1.95, 2.00  \\
\verb=v7.0_ML10_A176_MNXYZ=   &  3&0.417                  & 0.22             &$2\cdot10^{-11}$   &  10                             &    700                    &   1.76         &  1.85, 1.90, 1.95, 2.00  \\
\verb=v7.0_ML10_A177_MNXYZ=   & 3& 0.417                  & 0.22             &$2\cdot10^{-11}$   &  10                             &    700                    &   1.77         &  1.85, 1.90, 1.95, 2.00  \\
\\
\hline        
\hline                                   
\end{tabular*}
\label{tab:kla}
\begin{flushleft}
(a) Model name used as a reference;  \verb=XYZ=$=100*M_{X}$.  
(b) The triplet of input parameters CAK-$\alpha$, CAK-$k$, and density, $\rho_{0}$, at the surface of the donor star   in gr cm$^{-3}$. 
(c) The mass-loss rate in $10^{-7}\, M_{\odot}$ yr$^{-1}$. 
(d) The wind terminal velocity in km s$^{-1}$. 
(e) The binary separation in $R_{*}$. 
(f) The masses of the neutron star in $M_{\odot}$.
\end{flushleft}
\end{table*}

The wind acceleration depends on the CAK-$k$ and CAK-$\alpha$  parameters representing the number 
and strength of the absorption lines, and $\rho_{0}$, the  density
at the surface of the donor star.  Each of the triplet ($k$, $\alpha$, $\rho_{0}$)  produces a 
monotonically increasing wind velocity. Validation of the wind is performed 
by fitting the $\beta$-velocity law to the simulated data which returns the wind terminal velocity and 
the gradient of the velocity field. The  $\beta$ parameter  was found to be $\sim0.8$ in all cases and  
the terminal velocity ranged between 500 and 1200 km s$^{-1}$. 
The corresponding mass-loss rate (using the continuity equation) is ranging from  $2\times 10^{-7}\, M_{\odot}$ yr$^{-1}$ to 
$5\times 10^{-6}\, M_{\odot}$ yr$^{-1}$.
Table \ref{tab:kla} shows the CAK  parameters and the corresponding mass-loss rates and terminal velocities. 
 Although the stellar parameters are fixed in our simulations, 
observational uncertainties are
present. The simulated absorbing 
column density profile can be affected by these uncertainties (see
sect. \ref{sect:discuss_massns} for more details).

We used the above CAK parameters  to trigger the  2-D simulations. 
 The system now hosts a neutron star of mass $M_{X}$ (in $M_{\odot}$), having a luminosity $L_{X}=10^{36}$ erg s$^{-1}$ (MW11), 
 placed at a distance $\alpha$ (in $R_{*}$) from its donor. 
A  grid of parameters  with varying  $\alpha$ and M$_X$ was used to simulate the system. 
The parameters are listed in table \ref{tab:kla}.


\section{Results} \label{sec:results}

\subsection{Wind terminal velocity and mass-loss rate}

In order to study and constrain the dependency of the absorbing column density  on the wind properties, 
we built a grid of simulation for a variety of 
mass-loss rates and wind terminal velocities. 
We have used two wind terminal velocities  
(500 km s$^{-1}$ and 1200 km s$^{-1}$). 
and mass-loss rates of 
(0.2, 1, 5) $\times 10^{-6}\, M_{\odot}$ yr$^{-1}$. 
The orbital radius is fixed to $\alpha$=1.75 $R_{*}$ and the mass 
of the neutron star varies in the range 1.55, 1.75, and 1.95 $M_{\odot}$.
Six wind configurations (CAK-$\alpha$, CAK-$k$, $\rho_{0}$) were simulated  for each of these masses.
These parameters are listed in  table \ref{tab:kla}  (group 1)

The geometry of the shock, formed around the neutron star, is affected by the wind parameters
 and by the ionization.  The position and orientation of the bow shock  depends on the relative velocity of the neutron star within the wind 
and on the X-ray luminosity.

\begin{figure*}
  \centering
  
  \subfloat[$\upsilon_{\infty} \approx 500$ km s$^{-1}$ ; $\dot{M}_{w}\sim 2 \times 10^{-7}\,  M_{\odot}$ yr$^{-1}$]{\label{fig_IGRJ_ML2_v4}\includegraphics[width=0.42\textwidth]{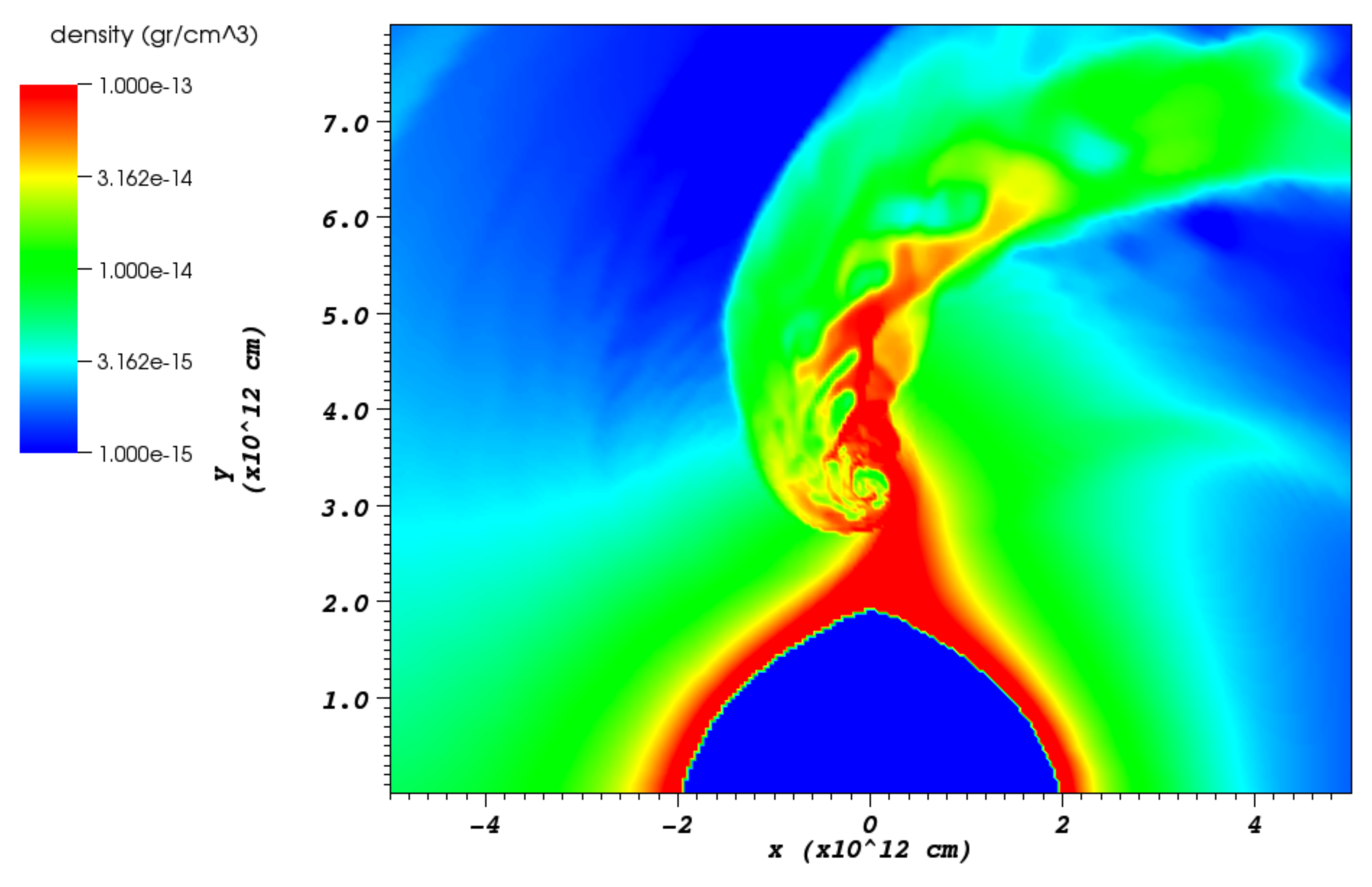} }                
  \subfloat[$\upsilon_{\infty} \approx 1200$ km s$^{-1}$ ; $\dot{M}_{w}\sim 2 \times 10^{-7}\, M_{\odot}$ yr$^{-1}$]{\label{fig_IGRJ_ML2_v12}\includegraphics[width=0.42\textwidth]{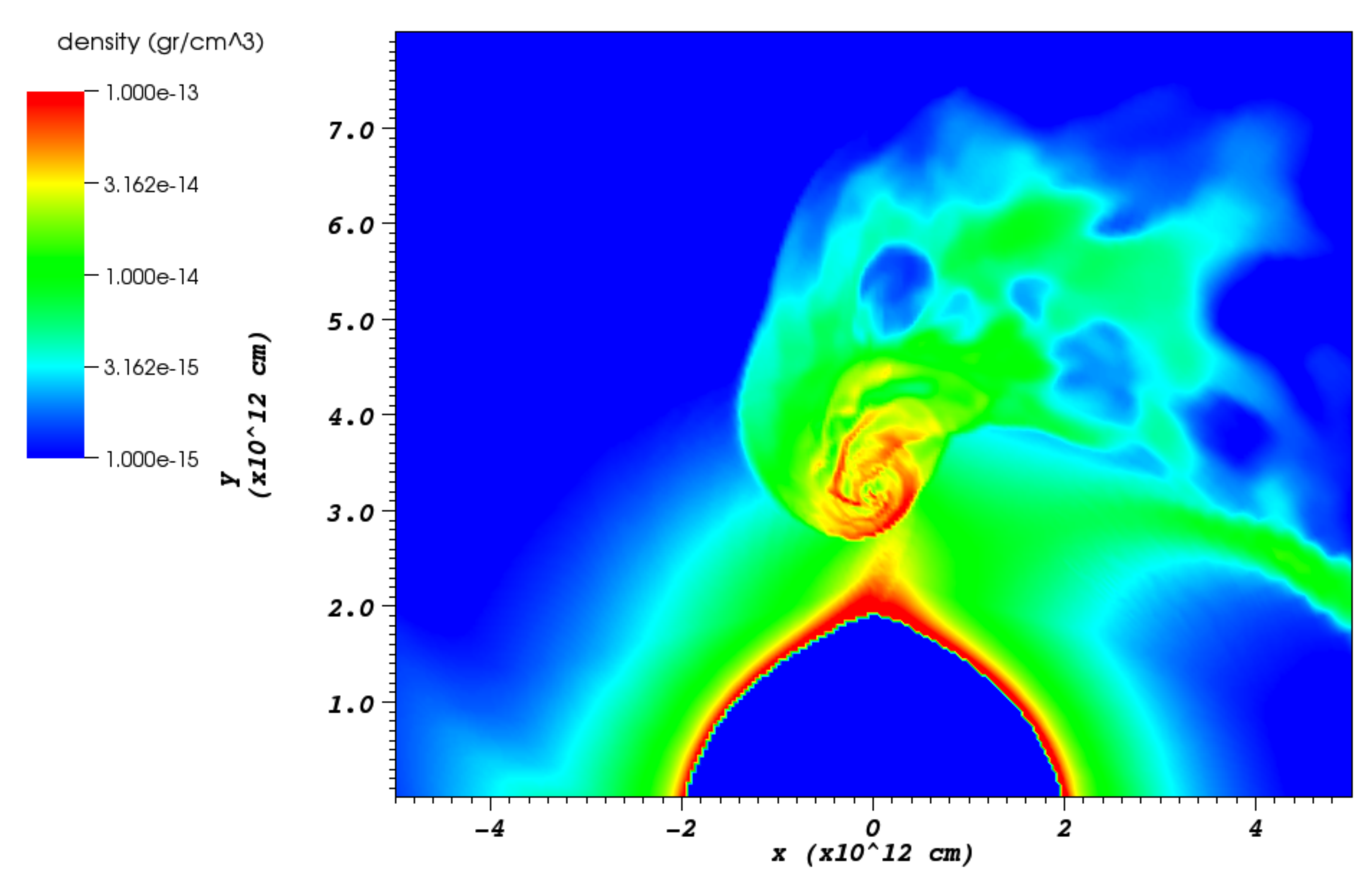}   }             
  
  \subfloat[$\upsilon_{\infty} \approx 500$ km s$^{-1}$ ; $\dot{M}_{w}\sim 1  \times 10^{-6}\, M_{\odot}$ yr$^{-1}$]{\label{fig_IGRJ_ML10_v4}\includegraphics[width=0.42\textwidth]{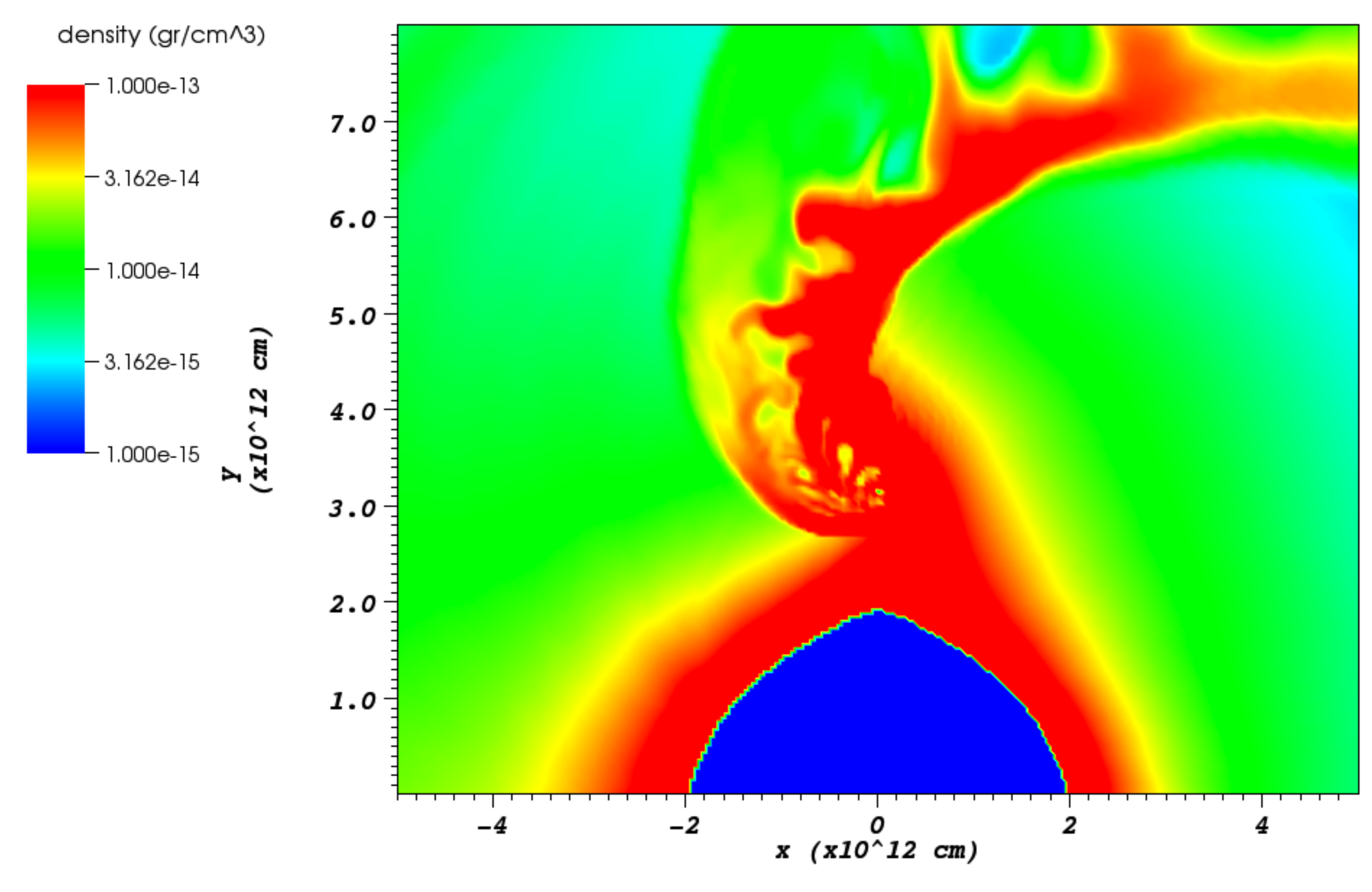} }               
  \subfloat[$\upsilon_{\infty} \approx 1200$ km s$^{-1}$ ; $\dot{M}_{w}\sim 1\times 10^{-6}\, M_{\odot}$ yr$^{-1}$]{\label{fig_IGRJ_ML10_v12}\includegraphics[width=0.42\textwidth]{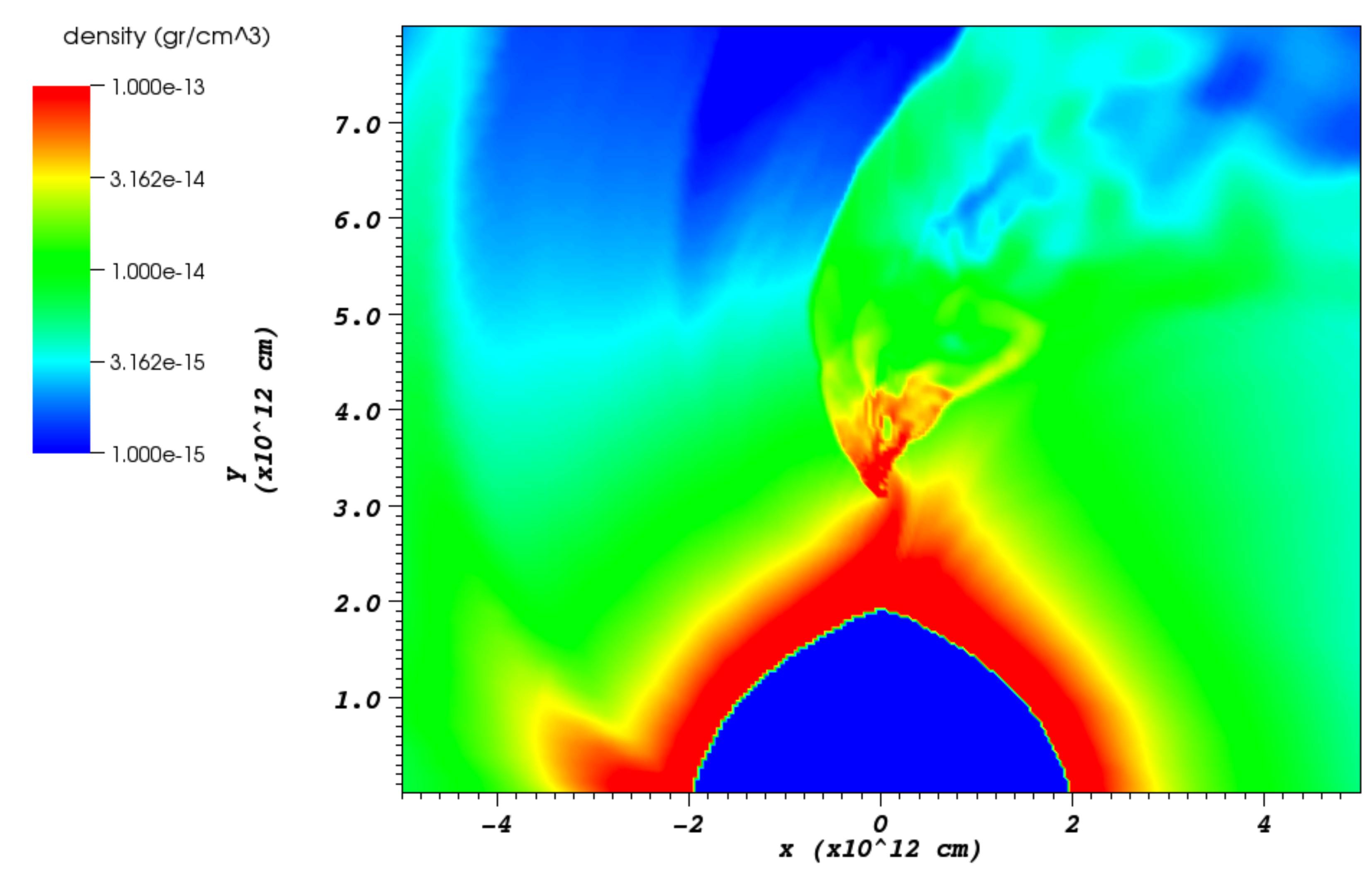}   }

  \caption{Density distribution (gr cm$^{-3}$; color bar) in the plane of the orbit after $\sim$ 3 orbits for 
  a neutron star mass of $M_{X}=1.95\, M_{\odot}$ and different wind parameters.  }
\label{fig:IGR_AlldotM}
\end{figure*}

Figure \ref{fig:IGR_AlldotM} shows the density maps (after 3 orbits) for some wind configurations. 
The orientation of the bow shock and  accretion wake (tail) depends on the mass-loss rate and wind terminal velocity.
For the slow wind, the  accretion wake is always visible, enhanced, and tilted.
For a wind velocity of 1200 km s$^{-1}$, the accretion wake is not that tilted. 
For a mass loss rate of $\dot{M}=2\times10^{-7}\, M_{\odot}$ yr$^{-1}$, 
the accretion wake is much weakened. 

\begin{figure}[!h]
\centering
\includegraphics[width=0.48\textwidth,angle=0]{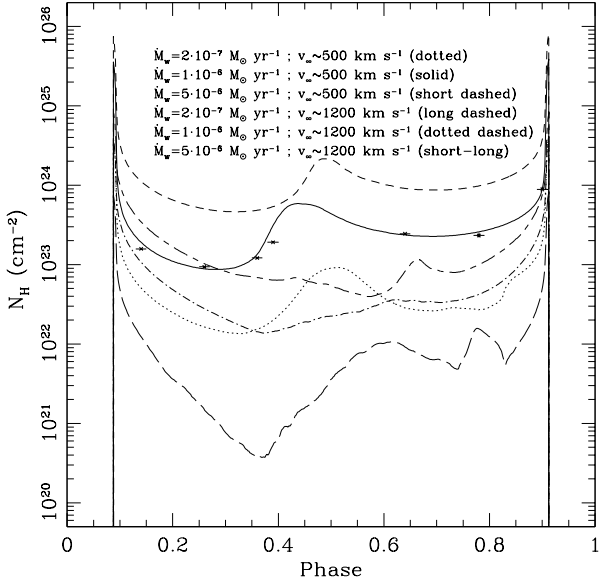} 
\caption{Time-averaged simulated absorbing column density ($N_{H}$) 
 (group 1 in table \ref{tab:kla}). 
The mass of the neutron star is fixed to M$_{X}$=1.95 $M_{\odot}$.
 The data points are from MW11.
  }
\label{fig:NH_MLR}
\end{figure}

Figure \ref{fig:NH_MLR} shows the time-averaged simulated absorbing column density as a
function of orbital phase, for M$_{X}$=1.95 $M_{\odot}$  
with the observations of IGR $J17252-3616$. 
  All these profiles show some 
 common characteristics. At early phases, the smooth wind component dominates, followed by a rise of $N_{H}$ (orbital 
 phase $\sim$ 0.3), representing the 
 bow shock and the accretion wake. Then, 
 the absorption roughly remains at a constant level. 
The  tidal stream between the donor and neutron star also influences the absorbing column density at later phases.

 Among these curves, only one simulation can represent the data
(\verb=v5.0_ML10_A175_MN195=) suggesting  a wind terminal velocity of $\upsilon_{\infty}=500$ km s$^{-1}$ and 
a mass loss rate of $\dot{M}=10^{-6}\, M_{\odot}$ yr$^{-1}$.
The dependency of  the absorbing column density orbital profile with the mass of the neutron star and binary separation parameters are discussed in 
sections \ref{sec:mass_ns} and \ref{sec:bsep}, respectively.

\subsection{Mass of the neutron star}
\label{sec:mass_ns}

A group of simulations was implemented  with neutron star masses varying between 1.5 and 2.0 $M_{\odot}$.
 The summary of these simulations is given in table \ref{tab:kla} (group 2). 
The separation  and the wind parameters are identical  to those of the model \verb=v5.0_ML10_A175_MN195=.

\begin{figure}[!h]
\centering
\includegraphics[width=0.45\textwidth,angle=0]{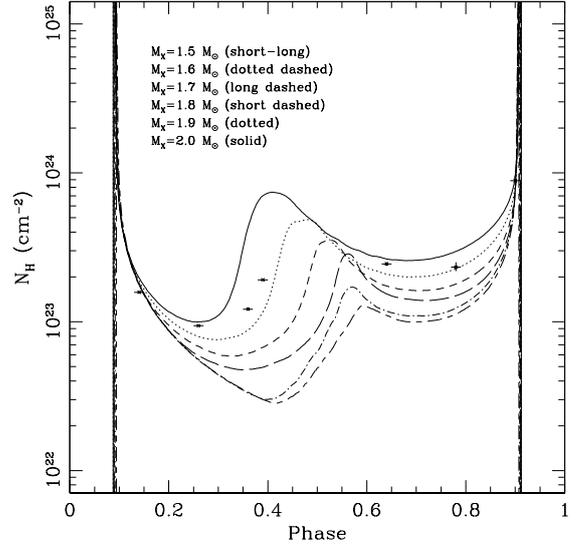} 
\caption{Time-averaged simulated absorbing column density ($N_{H}$) for various neutron star masses
(group 2 in table \ref{tab:kla}). The stellar wind is characterised by a 
$\dot{M}=1\times 10^{-6}$ M$_{\odot}$ yr$^{-1}$ and $\upsilon_{\infty}$=500 km s$^{-1}$. 
The data points are  from MW11. }
\label{fig:NH_1}
\end{figure}

\begin{figure*}
  \centering
  \subfloat[$M_{X}=1.5 M_{\odot}$]{\label{fig_IGRJ_Mx15}\includegraphics[width=0.42\textwidth]{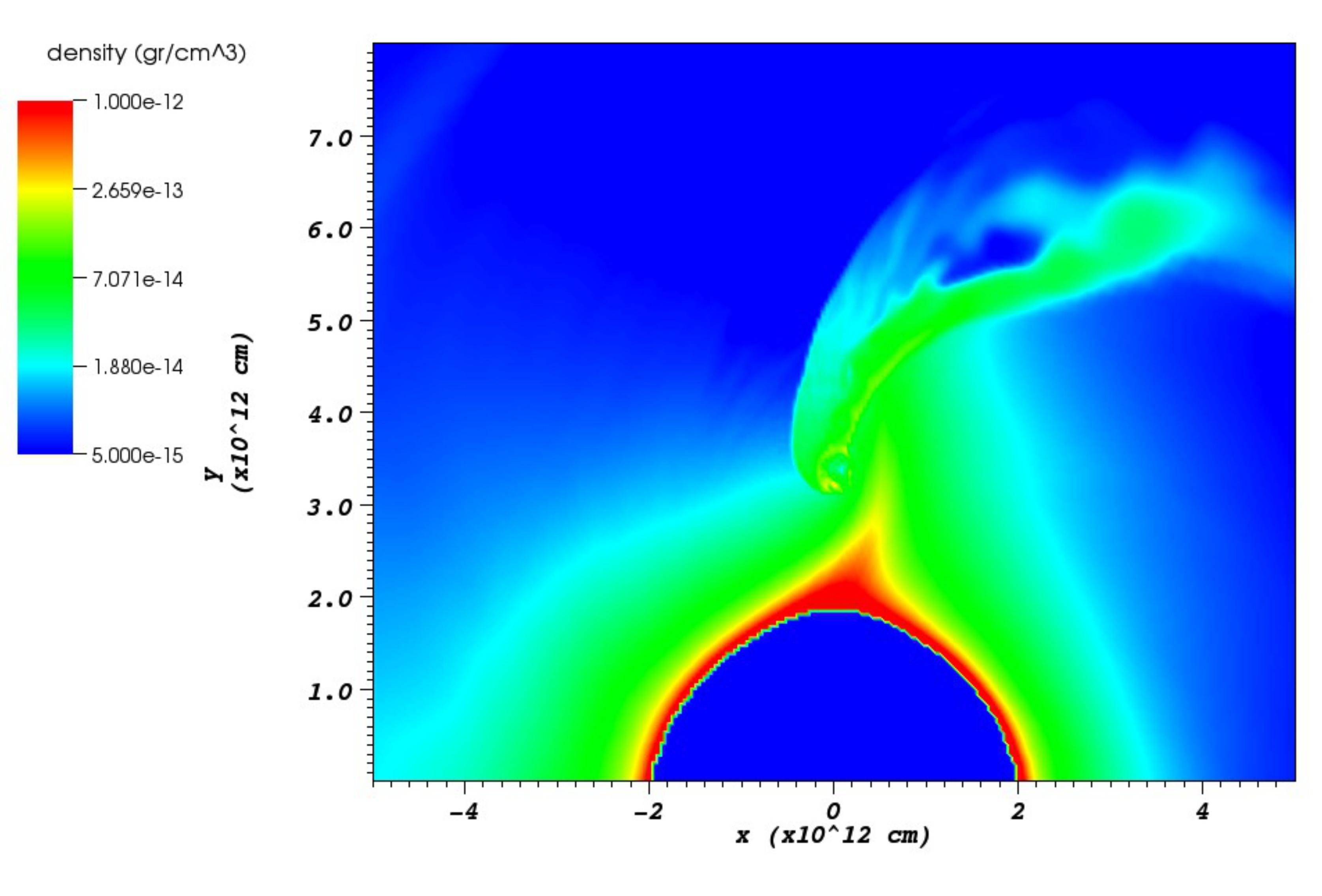}}                
  \subfloat[$M_{X}=1.7 M_{\odot}$]{\label{fig_IGRJ_Mx16}\includegraphics[width=0.42\textwidth]{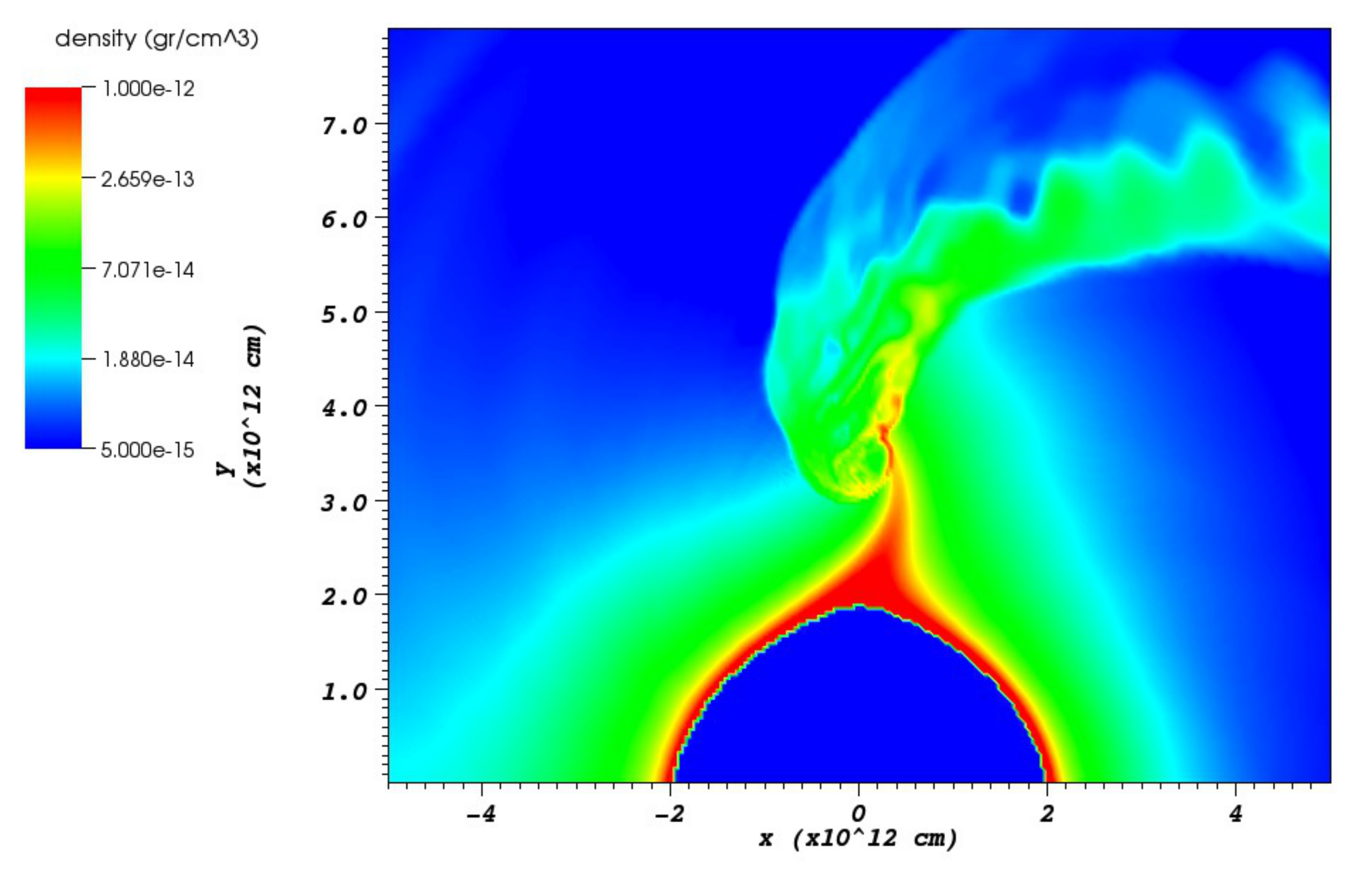}}

  \subfloat[$M_{X}=1.9 M_{\odot}$]{\label{fig_IGRJ_Mx19}\includegraphics[width=0.42\textwidth]{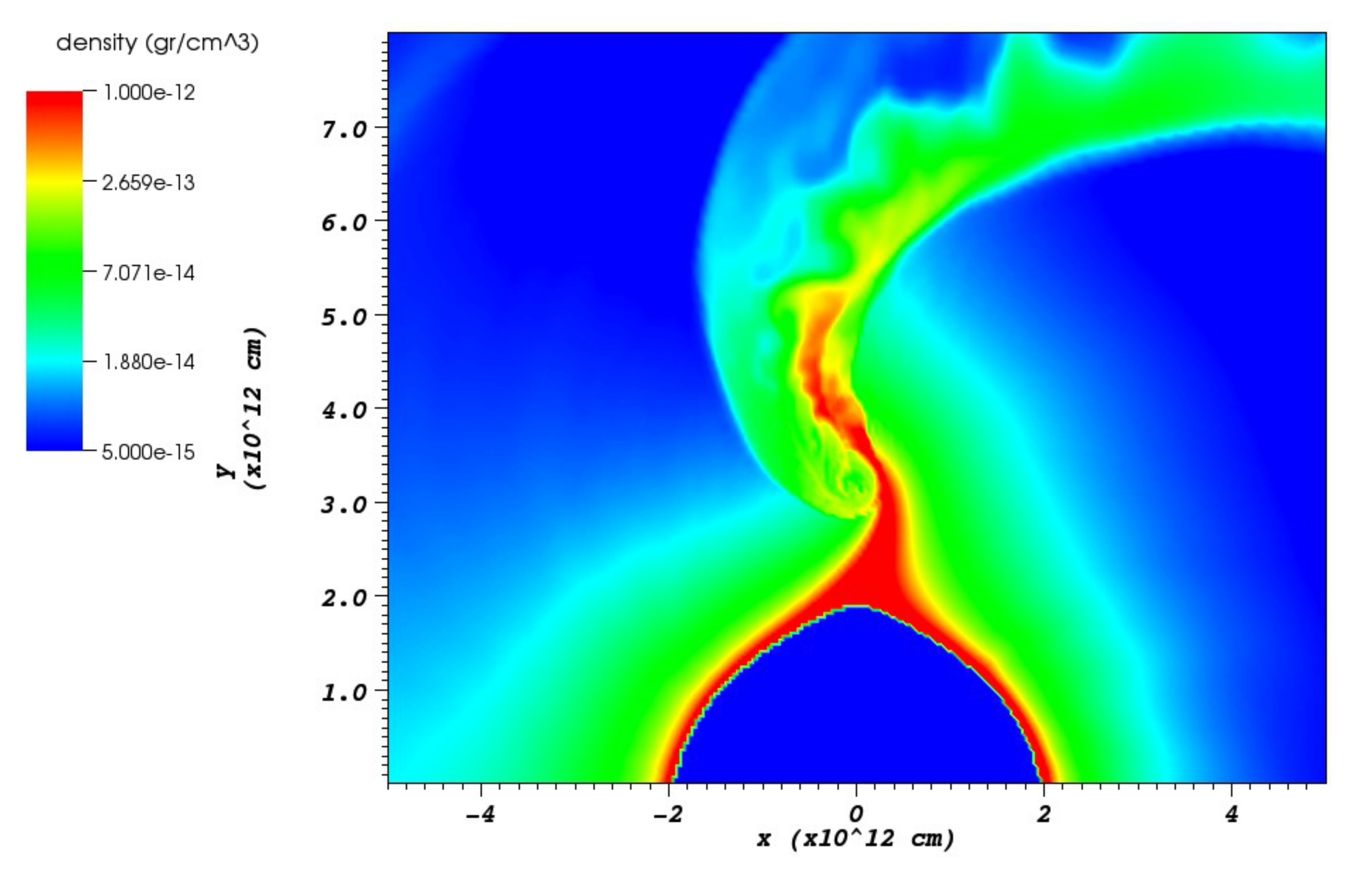}}                
  \subfloat[$M_{X}=2.0 M_{\odot}$]{\label{fig_IGRJ_Mx20}\includegraphics[width=0.42\textwidth]{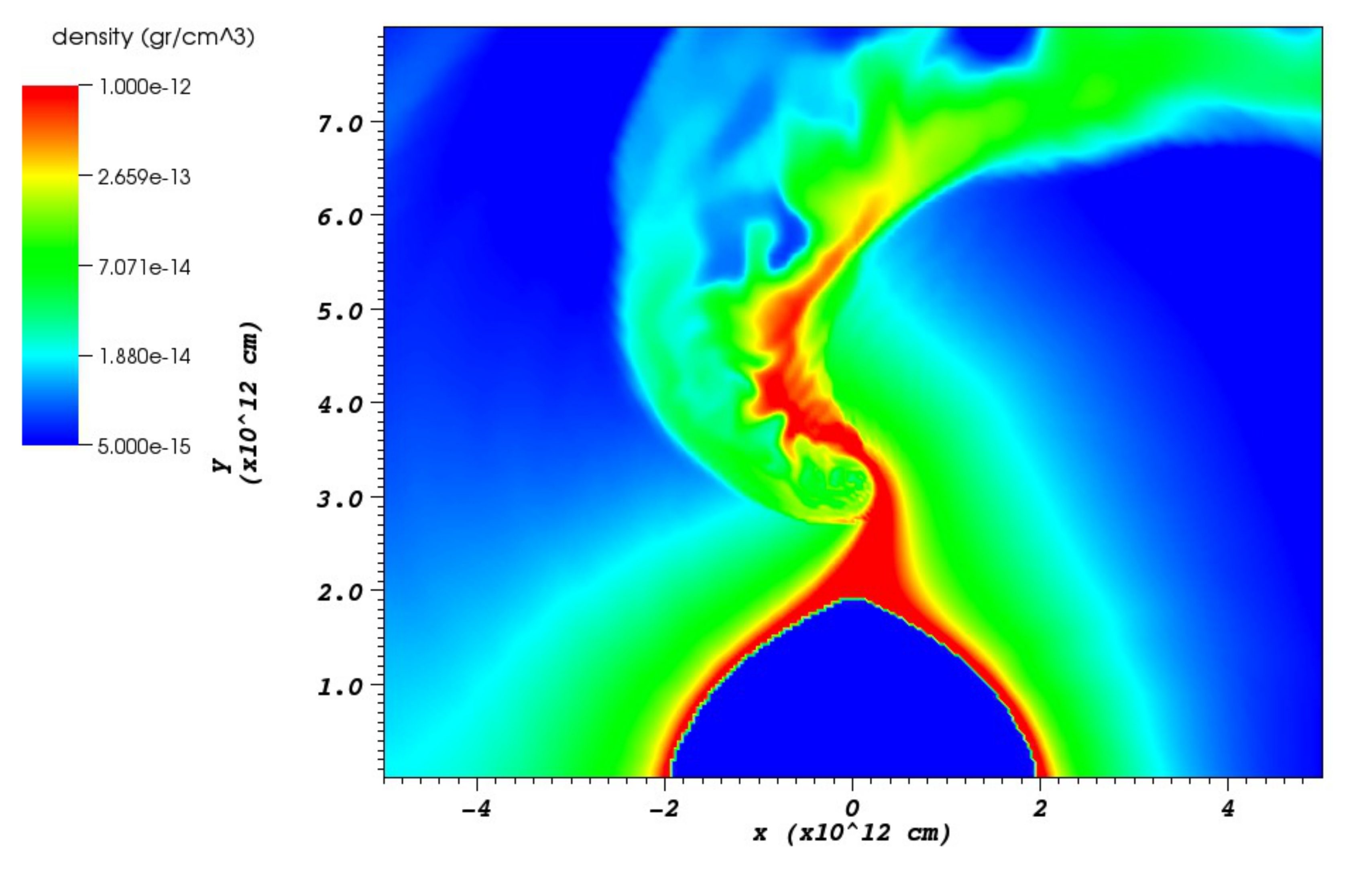}}                

  \caption{Density distribution (gr cm$^{-3}$; color bar) in the plane of the orbit after $\sim$ 3 orbits with a terminal velocity of $\upsilon_{\infty}\approx 500$ km s$^{-1}$ and a mass-loss rate $\dot{M}_{w}\approx 10^{-6}\, M_{\odot}$ yr$^{-1}$. The mass of the neutron star scales
  from 1.5 to 2.0 $M_{\odot}$. See the text for more details. }
\label{fig:IGR_AllM}
\end{figure*}

Figure \ref{fig:IGR_AllM} shows the density maps (after $\sim$ 3 orbits).  
The corresponding time-averaged absorbing column density is plotted in figure \ref{fig:NH_1}.
The  heavier  the neutron star, the more bended the accretion wake. This effect
is  revealed in the orbital phase dependency of the absorbing column density, moving the position of the  
minimum to earlier phases. For an heavier neutron star, more gas will accumulate and the  
absorption will get stronger.  
We can therefore estimate, in the frame of our model, that the mass of the neutron star  in in the range 
$M_{NS}=1.9-2.0\, M_{\odot}$ (see fig. \ref{fig:NH_1}). 

\subsection{Binary separation}
\label{sec:bsep}

The wind structure  is also  affected by the binary separation.
When the neutron star is closer to the donor star, the tidal stream between the two stars is enhanced.
 This effect is  maximized when the donor star fills its Roche lobe. In classical 
and obscured sgHMBs it is very likely that the evolved companion is close to filling its  Roche lobe. 
Therefore, tidal connection between the two stars enhances the absorbing column density \citep{Blondin91}.

\begin{figure*}[!ht]
\centering
\includegraphics[width=0.33\textwidth,angle=0]{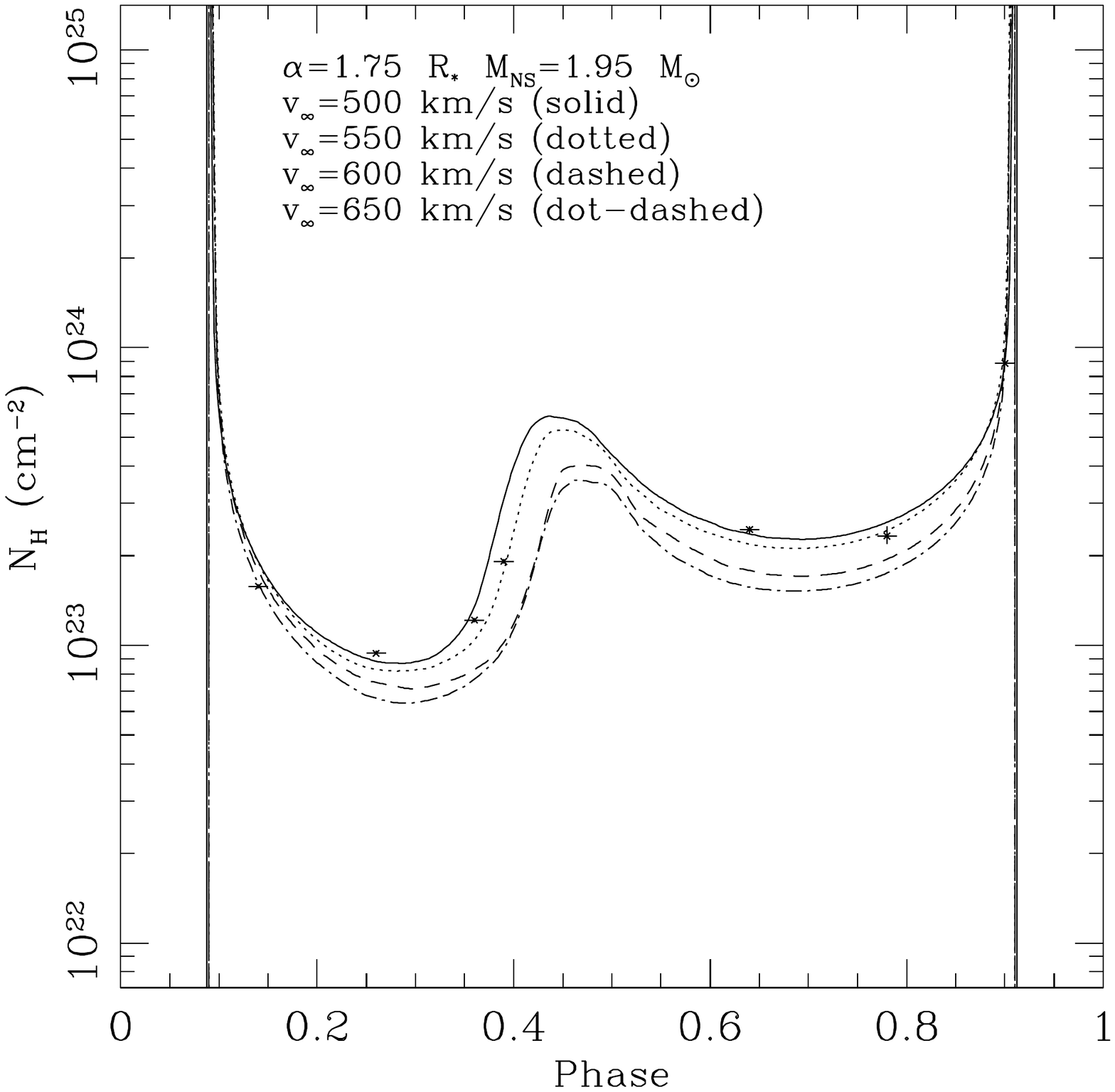} 
\includegraphics[width=0.33\textwidth,angle=0]{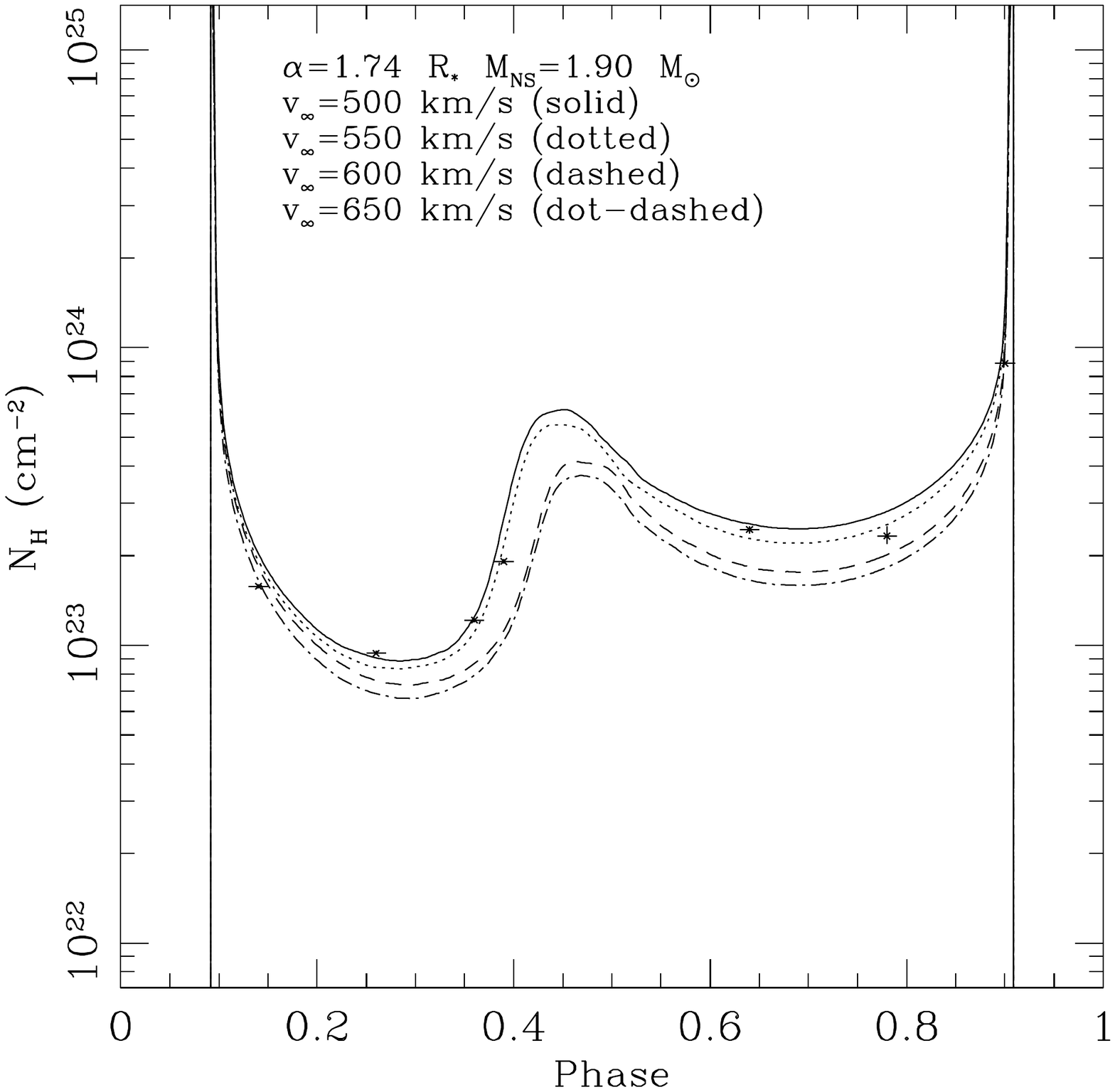} 
\includegraphics[width=0.33\textwidth,angle=0]{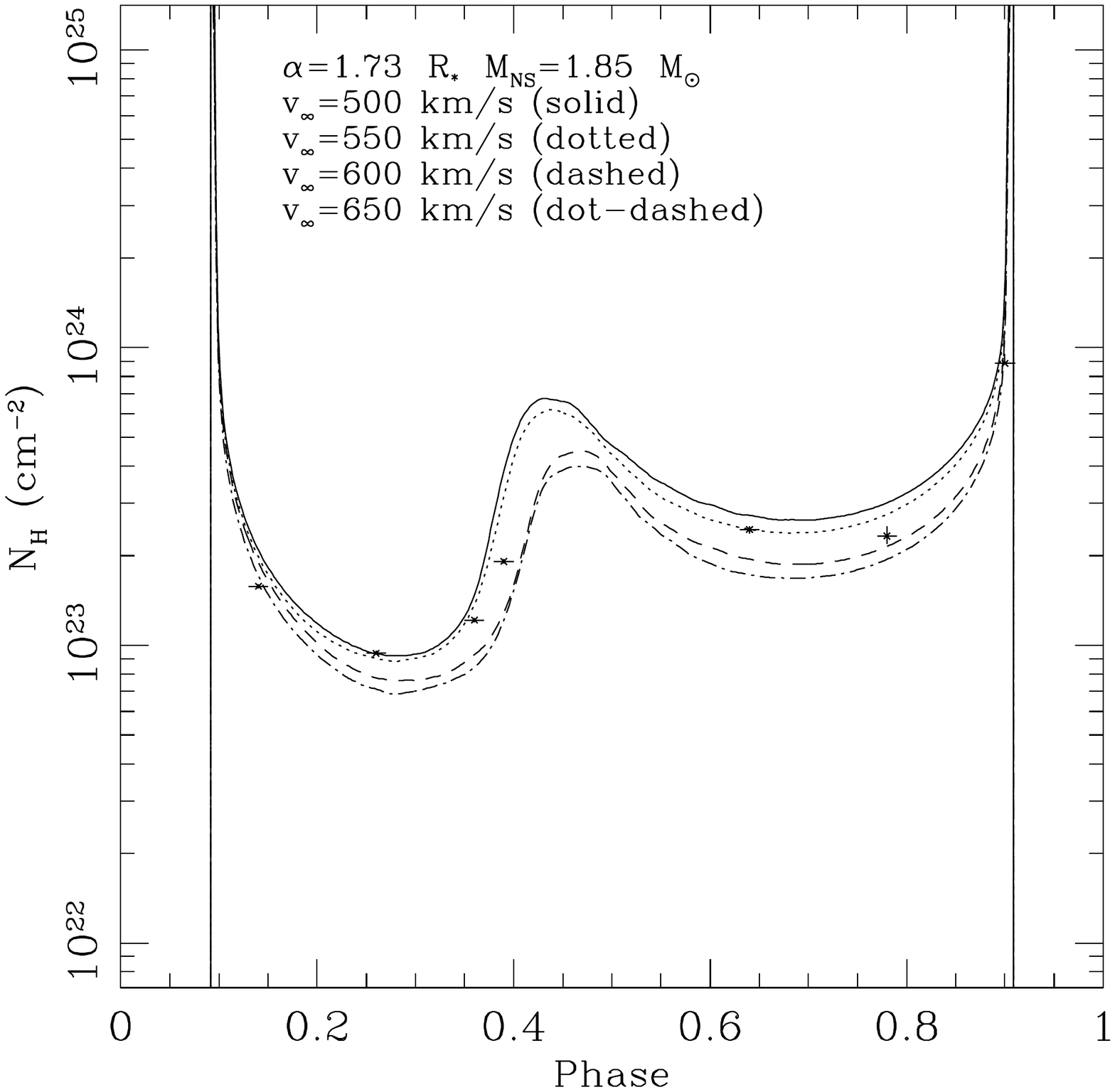} 
\caption{Time-averaged absorbing column density ($N_{H}$) for various parameters
(see group 3 in table \ref{tab:kla}). The stellar wind is characterized by 
$\dot{M}=1\times 10^{-6}\, M_{\odot}$ yr$^{-1}$  
The data points are from MW11. }
\label{fig:NH_sep_mass}
\end{figure*}

Figure \ref{fig:NH_sep_mass} shows the absorbing column density 
for various wind terminal velocities . In these plots
there is an indication that for a combination of neutron star mass and binary separation we sustain a roughly constant obscuration. 
Table \ref{tab:lognorm} gives the $\chi^{2}$ of the simulation (from fig. \ref{fig:NH_sep_mass}) accounting for both 
the error bars of the data and temporal variations of the absorbing column density. 
The  data can be represented by the model with a wind terminal velocity, neutron star mass, and binary separation in the range 
of 500-600 km s$^{-1}$, 1.85-1.95 $M_{\odot}$, and 1.74-1.75 $R_{*}$, respectively.

\begin{table}[!ht]
\caption{The $\chi^{2}$ for models for various neutron star masses, 
binary separations, and wind terminal velocities.}
\centering                          
\begin{tabular*}{0.35\textwidth}{@{\extracolsep{\fill}}  l   c  }
\hline
\hline                 
\\
Model name                                 &    $\chi^{2}$      \\
\\
\hline
\\
\verb=v5.0_ML10_A176_MN200=      &  3.5    \\

\verb=v5.0_ML10_A175_MN195=      &  0.9            \\
\verb=v5.0_ML10_A175_MN190=      &  1.8            \\
\verb=v5.0_ML10_A175_MN185=      &  1.9            \\

\verb=v5.0_ML10_A174_MN195=      &  1.3           \\
\verb=v5.0_ML10_A174_MN190=      &  0.8            \\
\verb=v5.0_ML10_A174_MN185=      &  1.1            \\

\verb=v5.0_ML10_A173_MN195=      &  1.5            \\
\verb=v5.0_ML10_A173_MN190=      &  2.3            \\
\verb=v5.0_ML10_A173_MN185=      &  2.8            \\

\hline
\\
\verb=v5.5_ML10_A175_MN195=         &  0.8        \\
\verb=v6.0_ML10_A175_MN195=       &   1.7          \\
\verb=v6.5_ML10_A175_MN195=          &  3.0       \\

\verb=v5.5_ML10_A174_MN190=      &  0.7\\
\verb=v6.0_ML10_A174_MN190=      &   1.3  \\
\verb=v6.5_ML10_A174_MN190=      &  1.9                 \\

\verb=v5.5_ML10_A173_MN185=      &  1.5   \\
\verb=v6.0_ML10_A173_MN185=      &   1.4    \\
\verb=v6.5_ML10_A173_MN185=    &  1.9        \\
\hline        
\hline                                   
\end{tabular*}
\label{tab:lognorm}
\end{table}

\subsection{Statistical analysis of the X-ray light-curve}

\SaveVerb{heavens}|http://www.isdc.unige.ch/heavens|

We have compared the  observed X-ray  light-curves of IGR $J17252-3616$ with these derived from our simulations. 
The observed light-curves of IGR J$17252-3616$ were obtained form the INTEGRAL \citep{integral-ref} soft $\gamma$-ray imager ISGRI  \citep{isgri-ref}
in the energy range 20-60 keV using the HEAVENS\footnote{\protect\UseVerb{heavens}} interface \citep{heavens-ref}.

Light-curves were built using data from the source and from a source free near-by region,
  using 1-hour, 3-hour, 6-hour and 12-hour  time-bins and flux histograms were constructed.  
 To achieve a relatively narrow width for the background fluctuations and  a good time resolution,  we selected the 6-hour time-bins for our analysis. 
 
\begin{figure*}
  \centering
  \includegraphics[width=0.55\textwidth]{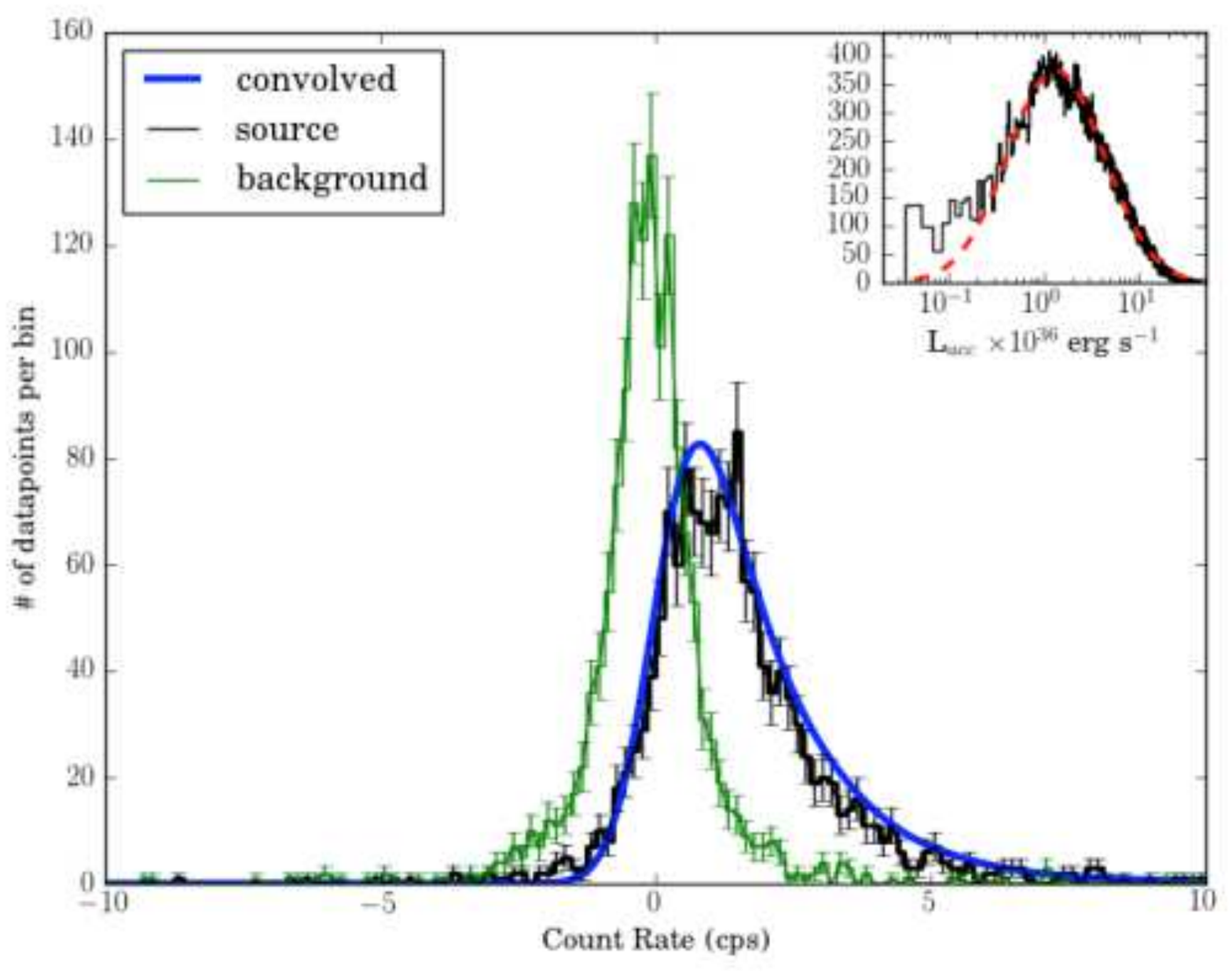}                
  \includegraphics[width=0.4\textwidth]{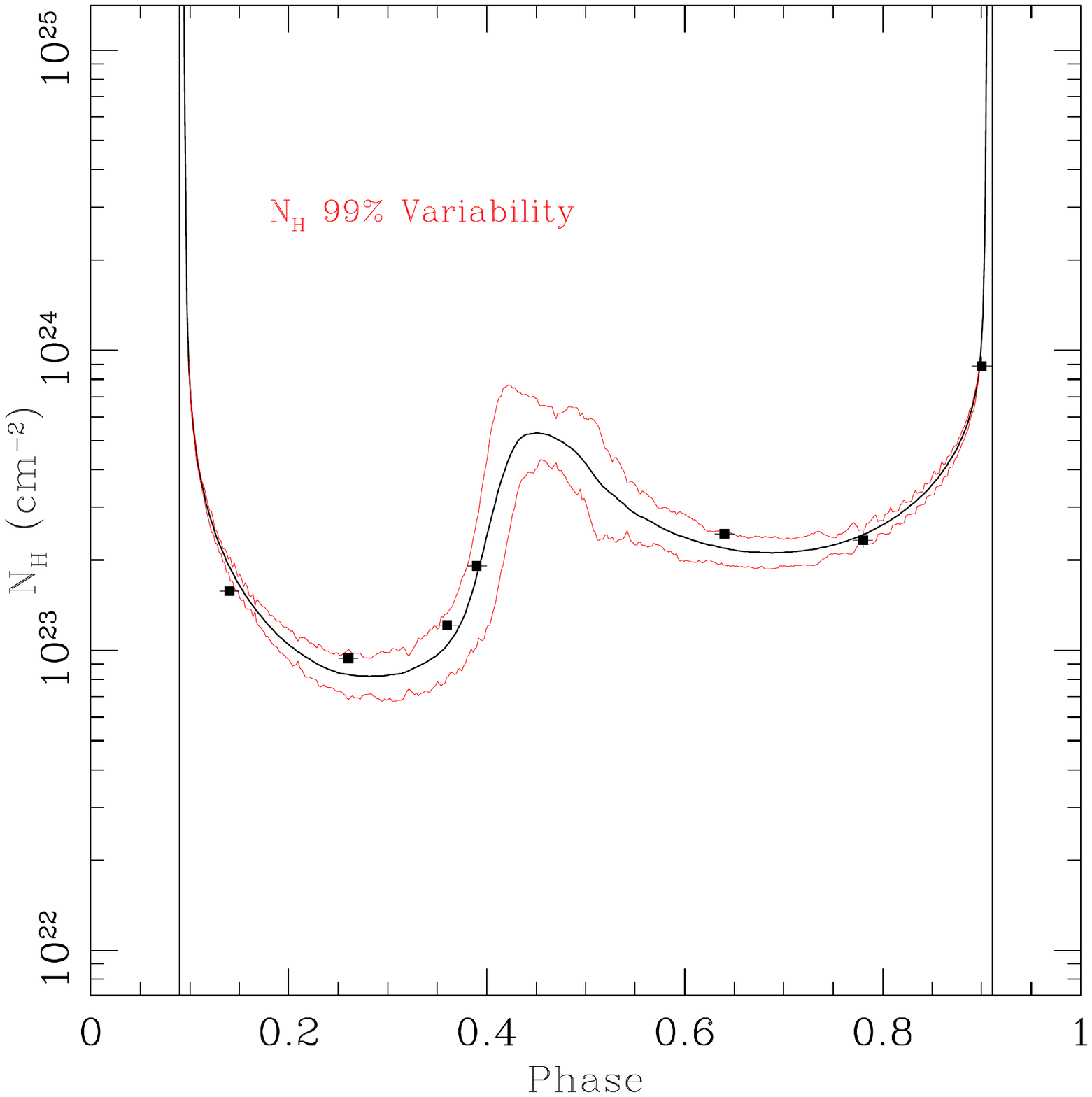}        
 
  \caption{Results from the simulation 
 \protect\UseVerb{bestmodel2}.  \emph{Left:} The histogram of observed light-curve  (black), the background fluctuation (green), and the convolution of the simulated 
 data with the latter (blue). 
The inset shows the corresponding distribution of the simulated. light-curve (black) with  a log-normal fit (red dashed curve) .  \emph{Right:} The corresponding 
  simulated time-averaged absorbing column density, together with its 99 \%  temporal variability, and the observed data from MW11.}
\label{fig:BESTSIM2}
\end{figure*}

The histogram of the observed fluxes can be fitted with  normal (gaussian) distributions.
The observations can be characterized by  a mean count rate of  $\mu$=1.1 cps and a width  $\sigma = 1.0$ cps for the source and 
 $\mu$=0 cps with a width $\sigma = 0.6$ cps  for the background (black and green lines in fig.  \ref{fig:BESTSIM2}).

We  performed a statistical analysis of  the simulated light-curves
which have  been built using the mass accretion rates derived from the 
hydrodynamical simulations,  translated  into  the instantaneous luminosity using $L_{acc}=\eta\dot{M}_{acc}c^{2}$, where $\eta=0.1$.  
 One should notice that the input parameter $L_{X}=10^{36}$ erg s$^{-1} = {\rm const.}$ (see sect. \ref{sec:code}) and the instantaneous
 X-ray luminosity are not the same. However, the average $L_{acc}$ is within the 10\% of the input parameter $L_{X}$. 
 The $L_{acc}$ histogram can be  fitted  by a log-normal distribution  (see inset  of fig. \ref{fig:BESTSIM2}). 
 The lognormal fit resulted in a luminosity $L_{X}\sim 1.1 \times 10^{36}$ erg s$^{-1}$ having a width of $\sigma=0.44$ for  model \verb=v5.5_ML10_A175_MN195=.

To compare the simulated distributions  with the observed ones, we  convolved 
the simulated data  with  the background fluctuations (blue curves in fig.  \ref{fig:BESTSIM2}).
We, also, plotted the simulated $N_H$ for this model together with 
its 99\% temporal variability  (right panel of fig. \ref{fig:BESTSIM2}).

\SaveVerb{bestmodel2}|v5.5_ML10_A175_MN195|

\section{Discussion}   \label{sec:discussion}

\subsection{Wind terminal velocities in obscured systems}

The wind terminal velocities of supergiant OB stars   is correlated with the effective temperature  
\citep[for a review see][]{winds_from_hot_stars}. For supergiant stars wind terminal velocities scales from
300 km s$^{-1}$ (for $T_{eff}$=10 kK) to 3000 km s$^{-1}$ (for $T_{eff}$=40 kK). An imprint of these outflows is 
 observed in the P-Cygni profiles of UV/Optical lines. For the system IGR J17252-3616, Optical/IR    
 studies  are  difficult due to high  interstellar absorption \citep{Rahoui_et_al08}.

For IGR J17252-3616, we suggested a wind terminal velocity of 500-600 km s$^{-1}$ which is 
about 2-3 times less than  expected for an OB supergiant of the same spectral type and effective temperature. 
The wind in massive binaries  can be slower than in isolated OB stars. The neutron star can indeed 
cut-off the acceleration by ionizing the local environment  \citep{1990ApJ...365..321S}.
Note that  the peculiar X-ray binary, GX 301-2, hosting 
a massive early type star (Wray 977), also features a low wind terminal velocity  of  $\upsilon_{\infty}\sim 500$ km s$^{-1}$ \citep{2006A&A...457..595K}. 
The column density in this system reaches  $\sim10^{24}$ cm$^{-2}$ and is 
 interpreted as the signature of a high density stream \citep{GXcolumndensity}. 
  In this context, it is also interesting to remark that some intermediately obscured sgHMXBs, such as $4U\, 1907+09$, feature intermediate wind terminal velocity in 
 the range $\sim 1000$ km s$^{-1}$ \citep{2010MNRAS.407.1182K}.
 
\subsection{ The mass of the neutron star}
\label{sect:discuss_massns}

The neutron star mass determined by the hydrodynamical simulations  is independent from the mass function 
derived from dynamical studies.  
However, it is  model-dependent and  uncertainties on the modelization  of the wind may have an impact.
 Figures \ref{fig:NH_1} and \ref{fig:NH_sep_mass} show that the mass of the neutron star is in the range  
$M_{X}=1.85-1.95\, M_{\odot}$.   This mass is  below the theoretical maximum mass of a neutron star, 
 $M_{NS}^{max}\lesssim$ 3 $M_{\odot}$ \citep{Muller1996508}.

The mass of the neutron star  and  the orbital radius are degenerate as  seen in fig.  \ref{fig:NH_sep_mass}. 
However, this degeneracy cannot allow very large or small masses of the neutron star (see table \ref{tab:lognorm}).  
The eclipse duration constrains the binary separation from  1.7 to 1.8 $R_{*}$ (MW11) and therefore the mass can scale from 1.85 to 2.0 $M_{\odot}$.
 A rather heavy neutron star is needed in any case to account for the observed profile.

 Radial velocity studies of the source, using VLT,  provided  estimates on  the 
 masses of the neutron star and donor star in the range $M_{NS}=1.46\pm0.38\, M_{\odot}$ and $M_{OB}=
 13.6\pm 1.6\, M_{\odot}$ for Roche lobe overflow;  $M_{NS}=1.63\pm0.38\, M_{\odot}$ and $M_{OB}=
 15.2\pm 1.6\, M_{\odot}$ for an edge-on view  \citep{Mason_et_al09b}. In both cases the mass of the neutron star 
 is consistent with our results
 
The companion radius and mass are obtained from the duration of the eclipse and stellar classification from mid-IR spectroscopy. These values 
are subject to observational uncertainties. 
To study the effects of these uncertainties on the determination of the mass of 
neutron star, we ran simulations of our best model for M$_{*}=15\pm 1\, M_{\odot}$ and R$_{*}=29 \pm 3\, R_{\odot}$. 
The effect of the stellar radius on $M_{NS}$ scales as 
$M_{NS}(R_{*})\propto 0.069\, R_{*}$. Given the range of acceptable stellar 
radius (i.e., 26 to 32),
the resulting neutron star's mass range varies from 1.75 $M_{\odot}$ to 2.15 $M_{\odot}$. 
The impact of the donor's mass on M$_{NS}$ scales as 
$M_{NS}(M_{*})\propto 0.13\, M_{*}$, resulting in, $M_{NS}=1.8-2.1\, M_{\odot}$.

Other supergiant HMXBs also feature large  neutron star masses.
 Vela X-1 hosts a neutron star with mass in the range  $M_{X}=1.88\pm0.13\, M_{\odot}$ \citep{Quaintrell_et_al03} 
 and   4U 1700-302 in the range $M_{X}=2.44\pm0.27\, M_{\odot}$ \citep{2002A&A...392..909C}. 
Although the sample is small, masses of neutron star in sgHMXBs, favors
 normal nucleonic EOSs  \citep{2007PhR...442..109L}. 
 

\section{Conclusions}   \label{sec:conclude}

In this paper we compare the observed properties with  the results of hydrodynamic simulations of 
the heavily obscured sgHMXB  IGR J17252-3616. 
Our conclusions are: 

\begin{itemize}

\item A low wind terminal velocity, $\upsilon_{\infty}=500-600$ km s$^{-1}$, is needed 
to explain the observations. A higher wind terminal velocity cannot produce enough obscuration to 
account for the observations. 

\item The neutron star mass 
can be constrained from the variability of the absorbing column density along the orbital phase. In the 
case of IGR J17252-3616  the mass of the neutron star is in the range 
1.75 $-$ 2.15 $M_{\odot}$.
 
\item The separation of the system can be limited in the range $\alpha$=1.73-1.75 $R_{*}$. 

\end{itemize}

Measuring the variability of the absorbing column density with the orbital phase in obscured 
sgHMXBs provides an independent constraint on  the mass of neutron stars.  



\bibliographystyle{aa} 
\bibliography{references} 

\end{document}